\documentclass[aps,prl,twocolumn,superscriptaddress,nofootinbib,nobalancelastpage,showpacs,nobibnotes]{revtex4}
\usepackage{epsfig}

\begin{document}
\title
{Bulk viscosity of a gas of neutrinos and coupled scalar particles, in the era
of recombination
}

\author{R. F. Sawyer}
\affiliation{Department of Physics, University of California at
Santa Barbara, Santa Barbara, California 93106}

\begin{abstract}
Bulk viscosity may serve to damp sound waves in a system of neutrinos coupled
to very light scalar particles, in the era after normal neutrino decoupling but before recombination.
We calculate the bulk viscosity parameter in a minimal scheme involving the
coupling of the two systems. We add some remarks on the bulk viscosity of
a system of fully ionized hydrogen plus photons.

\end{abstract}
\maketitle
\section{Introduction}
It has long been known that a simple form of bulk
viscosity is a potential source of dissipation in mixtures of relativistic
and nonrelativistic fluids. For the cosmologically interesting case of
photons and matter in the era just before recombination, where shear viscosity
does indeed strongly damp waves for wave-lengths up to some critical length,
Weinberg \cite{weinberg} has given the expression for the bulk viscosity parameter,
\begin{equation}
\zeta=4aT^4\tau_\gamma\Bigr[ {1\over 3}-\Bigr ({\partial P \over \partial \rho}\Bigr )_n \Bigr ]^2\,.
\label{weinberg}
\end{equation}
Here $\tau_\gamma$ is a mean free time for photon in the medium and the electron (and proton) number density $n$ is held fixed in the derivative. Since the
derivative  in (\ref{weinberg}) is of the order $(1/3-10^{-10})$ in the early universe, in view of the large
photon/electron ratio, 
the effect is absolutely negligible in this system. 

For what comes later in this paper, we
comment on the parameter, $\tau_\gamma$.
The development that led to (\ref{weinberg}) was based in part
on the solution by Thomas \cite{thomas} of a relativistic transport 
equation for photons. However, Thomas'
treatment included only photon absorption and emission. When Compton scattering is 
the interaction mechanism, it appears that $\tau_\gamma$ should be the mean free time for
Compton scattering weighted by an efficiency factor for energy transfer, that is to say,
multiplied by a number of order of the number of collisions required for
temperature equilibration of a system which initially has different temperatures
for the matter and for the radiation. When the scattering mechanism is Compton scattering
and the system is near the recombination temperature this is a large number indeed.

The motivation for the present paper is the possible existence of a scalar field, $\phi$,
with mass smaller than neutrino masses and coupled to neutrinos. Over time,
theorists have found a number of reasons to entertain such a possibility \cite{reas1}-
\cite{reas7}.
The most stringent
bounds on the coupling strength, at least until recently, have come from supernovae \cite{lim1}-
\cite{lim5}.
Recently Beacom, Bell and Dodleson \cite{beacom} demonstrated the rather surprising result that, staying
within these bounds,
processes $\nu+\nu\rightarrow \phi+\phi$ can remove most of the neutrinos
from the universe as the temperature declines into the region in which
the neutrinos become non-relativistic, a region that roughly corresponds
to the recombination region itself.
However Hannestad \cite{hanne} pointed out that the acoustic oscillations of the fluid
of interacting neutrinos and $\phi$'s in this picture could alter the CMB angular 
power spectrum in ways that conflict with the data \footnote{In ref. \cite{silk} from 1987 it had
already been noted that an interacting sea of neutrinos and majorons could form structure
in the universe; of course the details of the interesting application of these ideas
are now different, in view of all of the data that
has come in since then.}.

Beginning from this observation, Raffelt and Hannestad \cite{raff} have given coupling constant limitations 
that a true free-streaming requirement would put on the $\nu-\phi$ coupling.
But more detailed calculations of the effect of $\nu$ scattering, annihilation or decay 
on the CMB and the matter power spectrum appear to show that the landscape
of coupling constant limitations is considerably more complicated that dictated
by the ``free-streaming" criterion, and, indeed, that in some coupling
schemes, parameters that give a ``tightly coupled" $\nu -\phi$ fluid
at the recombination temperature are compatible with the data \cite{bell}.

The first purpose of the present paper is to point out a possible role for bulk
viscosity in controlling the size of neutrino density fluctuations under some
conditions. Clearly the region of application would be in between the
tightly coupled case and the free streaming case.
The origin of this dissipation is qualitatively similar to
that which underlies (\ref{weinberg}). We have two gases, one completely
relativistic, the other in the relativistic-non-relativistic transitional
region. Thus if we apply an adiabatic compression, they change temperature
by different amounts. There is a time lag in the equilibration, and consequent
creation of entropy. Now, however, the analog of the factor that produced
the square of the electron-photon ratio in (\ref{weinberg}) need not give
an extremely small number, since the numbers of $\nu$'s and $\phi$'s are
comparable.

In principle, a formalism for the evolution of inhomogeneities
that followed the detailed evolution of the position and momentum distributions
of each kind of particle would have no need at all to mention viscosity explicitly.
Indeed, shear viscosity is subsumed within the standard 
evolution codes for the baryon, electron, photon systems, entering at the quadrupole level
where non-isotropic stresses are taken into account. Bulk viscosity is not included, nor need 
it be, in the system of matter plus photons; but interacting neutrinos can
be another story.
 
There are two important ways in which our $\nu-\phi$ system requires 
computational mechanics somewhat different from those necessary to derive the Weinberg formula:

1) In contrast to the case with matter and photons, where there is a conserved number
of protons and electrons, the neutrino system will be taken to 
be symmetrical between particles and antiparticles, and the annihilation process
into $\phi$'s is an integral part of the evolution. This makes the determination of the 
analog of the factor in (\ref{weinberg}), 
$({1\over 3}-[{\partial p \over \partial \rho}\Bigr ]_n )$,
different than in the photon-matter case.

2) Looking at the parameters for the problem at hand it appears that the frequency
dependence of the bulk viscosity will play a role in the application. The
consequences of this dependence have been important in the role of bulk viscosity
in damping the gravitational radiation instabilities in rapidly rotating neutron stars \cite{bv1}-
\cite{bv4}.
This dependence, being essentially a non-local effect, does not rigorously
fit into the framework that is best suited to the analysis of a relativistic fluid, {\it viz},
addition of an additional term in the energy-momentum tensor in the comoving system,
\begin{equation}
\delta T^{i,j}=-\delta^{i,j} \zeta (\omega) \nabla \cdot {\bf U}\, ,
\label{stress}
\end{equation} 
where $\bf U$ is the space component of the four-velocity.
We believe that it is legitimate, however, to crank the formalism
forward using (\ref{stress}), and then, at the point when Fourier transforms
are introduced in position space, to evaluate $\zeta (\omega)$ at $\omega=v_s k$
where $v_s$ is the speed of sound in the medium.

In one respect, however, our $\nu-\phi$ system is simpler than the matter-photon
system; the energy equilibration in a single collision is relatively efficient. This is
because we have the direct transformation, $\nu+\nu \rightarrow \phi +\phi$.\footnote{
In models that couple light scalars to neutrinos, the neutrinos might be taken either
as Dirac neutrinos, with possible couplings that preserve lepton number, leading to
$\nu+\bar \nu \rightarrow  \phi +\phi$,
or as Majorana neutrinos that violate lepton number, in which case we have 
$\nu+\nu \rightarrow \phi +\phi$. The amplitudes for the two cases are different
(in angular dependence, for example) but we can take an estimate of total 
rate, as a function of coupling, $G$, to be the same in both cases. The distinction
will not enter further in this paper, except here and there in terminology.}

The formalism that we shall present is completely applicable to the 
case of photons and ordinary matter 
and this may possibly be of interest in other astronomical contexts.
We shall apply our method to the ionized-hydrogen system, with a result related to
(\ref{weinberg})
 
\section{Calculation}
We consider an adiabatic compression specified by a value of $-\delta V / V$, where we do not
allow energy transfer between the two species, leading to respective temperature changes for
the species of  $\delta T_{\phi,\nu}$. Defining the indices  $\gamma_{\phi,\nu}$ for the respective
species as,
\begin{eqnarray}
{\delta T_{\phi,\nu} \over T_{\phi,\nu}}= - {\delta V \over V} \Bigr (\gamma _{\phi,\nu} -1 \Bigr)\,,
\end{eqnarray}
we take 
\begin{eqnarray}
{ \delta V \over  V}=b e^{i \omega t}\,,
\end{eqnarray}
where it is to be understood that we consistently take real parts to get linear perturbations
in the physical quantities. 
The time rate of change of the temperatures is thus,
\begin{eqnarray}
{d \over dt} T_{\phi,\nu} =i b (\gamma_{\phi,\nu}-1) \omega T_{\phi,\nu} \,. 
\label{gammas}
\end{eqnarray} 
The rate of heat transfer per unit volume between the two baths is of the order of,
\begin{equation}
\dot Q= \Gamma \rho_\phi {(T_\phi-T_\nu)\over T_0} \,,
\label{heat}
\end{equation}
where $T_0$ is the mean temperature of both baths,
$\rho_\phi$ the energy density of the $\phi$ particle bath, and $\Gamma$ 
an effective collision rate for $\phi$'s in the medium. By ``effective" we
mean the rate for collisions of a $\phi$ with the $\nu$ cloud,
supplemented by rate for $\phi+\phi\rightarrow 2\, \nu$, and multiplied by an energy transfer efficiency factor which is of the order of, but less than, unity,  
(and model dependent as well). 
Adding the effect of heat transfer to the right hand side of (\ref{gammas}) gives,
\begin{eqnarray}
{d \over dt} T_{\phi}(t) =i (\gamma_{\phi}-1) b \omega T_0 e^{i \omega t}
-  \Gamma \rho_\phi
\Bigr [ {T_\phi(t)-T_\nu (t) \over T_0}\Bigr ][c_V^{(\phi)}]^{-1},
\nonumber\\
 {d \over dt} T_{\nu}(t) =i (\gamma_{\nu}-1) b \omega T_0
e^{i \omega t} + \Gamma \rho_\phi 
\Bigr[{T_\phi(t)-T_\nu (t)\over T_0}\Bigr ][c_V^{(\nu)}]^{-1}.
\nonumber\\
\,
\label{gammas2}
\end{eqnarray} 
Subtracting, defining $\delta T(t)=T_\phi (t)-T_\nu(t)$, and solving, we obtain,
\begin{eqnarray}
\delta T(t)={\rm Re} \Bigr[ {i \omega (\gamma_\phi-\gamma_\nu)T_0 b \over i\omega+ \Gamma
\rho_\phi([c_V^\phi]^{-1} +[c_V^\nu ]^{-1})T_0^{-1}} e^{i\omega t} \Big ]\,.
\label{temp-diff}
\end{eqnarray}
In the calculation of work, $\int P \,dV$, below, the volume differential will
be given by $[d/dt(\delta V)] dt=bV\omega \sin (\omega t)$. We will therefore retain only
the $\sin (\omega t)$ terms in the $\delta T$'s, since the $\cos (\omega t)$ terms
do not contribute to the dissipation. We denote these parts and the associated pressure
changes by $\delta T^s$, $\delta P^s$. 

From (\ref{gammas2}) we then find that the $\sin (\omega t)$ term in the oscillation of the temperature difference is divided
between the two baths as,
\begin{eqnarray}
\delta T^s_\phi(t)=\delta T^s(t) {[c_V^{(\phi)}]^{-1} \over [c_V^{(\phi)}]^{-1}+[c_V^{(\nu)}]^{-1}}\,,
\nonumber\\
\,
\nonumber\\
\delta T^s_\nu(t)=-\delta T^s(t) {[c_V^{(\nu)}]^{-1} \over [c_V^{(\phi)}]^{-1}+[c_V^{(\nu)}]^{-1}}\,.
\end{eqnarray}
Next we calculate the induced pressure changes of the two clouds,
\begin{eqnarray}
\delta P^s_\phi(t)=\delta T^s_\phi (t) \Bigl [{dP_\phi  \over dT}\Bigr ] \equiv \alpha_\phi \delta T^s_\phi(t) \,,
\nonumber\\
\,
\nonumber\\
\delta P^s_\nu(t)=\delta T^s_\nu (t) \Bigl [{dP_\nu  \over dT}\Bigr ] \equiv \alpha_\nu \delta T^s_\nu(t)\,.
\end{eqnarray}
Taking the $\phi$ mass to be zero, we list the three relevant parameters for the $\phi$ cloud, 
\begin{equation}
\alpha_\phi={2 \pi^2 T_0^3 \over 45}~,~c_V^{(\phi)}= 3 \alpha_\phi~,~ \gamma_\phi= 4/3 \,,
\label{params}
\end{equation}
where we use units $\hbar=c=k_B=1$.
For the neutrino cloud in the domain in which the mass plays a role, numerical calculations
of the corresponding parameters are described in the appendix. 

We write the net pressure oscillation amplitude as,
$\delta P^s_\phi+ \delta P^s_\nu$, obtaining,
\begin{eqnarray}
\delta P^s \equiv \delta P^s_\phi+\delta P^s_\nu =
 -\omega b \zeta \sin \omega t \,,
\nonumber\\
\,
\end{eqnarray}
where
\begin{eqnarray}
\zeta={(\gamma_\nu-\gamma_\phi)\Gamma \rho_\phi
\over \omega^2 + \Gamma^2 \rho_\phi^2( [c_V^{(\phi)}]^{-1}+[c_V^{(\nu)}]^{-1})^2 T_0^{-2}\, } 
\Bigl  [{\alpha_\nu \over c_V^{(\nu)}}-{1 \over 3} \Bigr]\, .
\nonumber\\
\label{bv0}
\,
\end{eqnarray}
The amount of mechanical energy converted to thermal energy in one period
is given by 
\begin{eqnarray}
\Delta E=\int_0^{2 \pi/\omega}dt 2 \zeta \omega \sin (\omega t) dV(t)= 2\pi \omega  b^2 \zeta \,,
\end{eqnarray}
giving the time averaged dissipation per unit volume,

\begin{eqnarray}
\dot  E/V=  \zeta \omega^2 b^2 /2 \,,
\end{eqnarray}
identifying $\zeta$ as the bulk viscosity.

Using eq.(5) of ref.\cite{bell} and staying in the temperature region that is (nearly) non-relativistic for the $\nu$'s we calculate the reaction rate for $\phi+\phi \rightarrow \nu +\nu$ in the  mixture as $\Gamma=G^4 T g(m_\nu/T)$, 
where $G$ is the $\nu -\chi$ coupling constant, and the dimensionless function $g(m_\nu/T)$ is given in the appendix.
Introducing $k\approx 3 \omega$, we write $\zeta$ in the form,
\begin{equation}
\zeta={10^{4}G^{-4} \rho_\phi T_0^{-1} H\Bigr({m_\nu \over T_0  }\Bigr )\over 1+k^2/k_0^2}\,,
\label{bv1}
\end{equation}
where
\begin{eqnarray}
H={10^{-4}(\gamma_\nu-1/3)\Bigr ({\alpha_\nu \over c_V^{(\nu)}}-1/3\Bigr )T_0^2 \over  
g(m_\nu/T_0)~\Bigr ([c_V^{(\phi)}]^{-1}+ [c_V^{(\nu)}]^{-1}\Bigr ) ^2  \rho_\phi^2}.
\end{eqnarray}
The ``cutoff" wave number $k_0$ is given as,
\begin{eqnarray}
     k_0=10^{-3} G^4 T_0 F\Bigr({m_\nu \over T_0  }\Bigr)\,,
\label{wnc}
\end{eqnarray}
where 
\begin{eqnarray}
F\Bigr({m_\nu \over T_0  }\Bigr)=10^3\,[g(m_\nu/T_0)]^{1/2}3^{-1/2}
\nonumber\\
\times \rho_\phi T_0^{-1}\Bigr ([c_V^{(\phi)}]^{-1}+ [c_V^{(\nu)}]^{-1}\Bigr ).
\nonumber\\
\,
\end{eqnarray}
$F$ and $H$ are dimensionless functions of $m_\nu /T_0$, plotted in figs. 1 and 2 for the range of
values in which we are most interested. They are both of order unity
in the semirelativistic region in which there may be significant effects of bulk viscosity.
The calculation
of the components of the above functions, $\alpha_\nu$, $\gamma_\nu$, and $g$ is
discussed in the appendix.  
\begin {figure}[ht]
    \begin{center}
       \epsfxsize 2.75in
        \begin{tabular}{rc}
           \vbox{\hbox{
$\displaystyle{ \, { } }$
               \hskip -0.1in \null} 
} &          \epsfbox{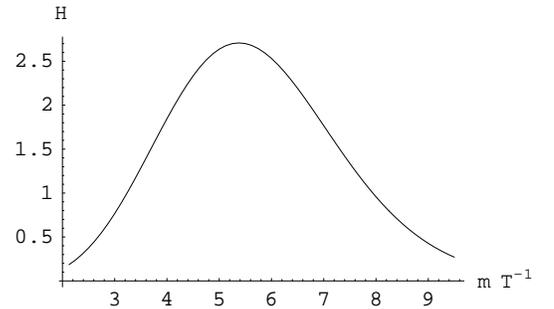} \\
            &
            \hbox{} \\
        \end{tabular}
    \end{center}
\label{fig.1}
\protect\caption
    {%
The function $H(m_\nu/T)$ which appears in the bulk viscosity result, (\ref{bv1}).
 }
\end {figure}

\begin {figure}[ht]
    \begin{center}
       \epsfxsize 2.75in
        \begin{tabular}{rc}
           \vbox{\hbox{
$\displaystyle{ \, { } }$
               \hskip -0.1in \null} 
} &
            \epsfbox{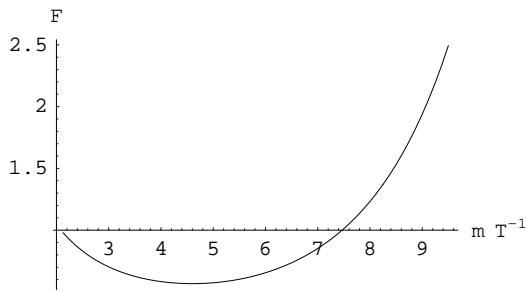} \\
            &
            \hbox{} \\
        \end{tabular}
    \end{center}
\label{fig.2}
\protect\caption
    {%
The function $F(m_\nu/T)$ which appears in the result for the wave-number cutoff $k_0$ (\ref{wnc}).
 }
\end {figure}
The damping rate for a plane sound wave in a medium with bulk viscosity, $\zeta$, is given 
by eq.(2.56) of ref.{\cite{weinberg},
\begin{equation}
\Gamma_{\rm{damp}}={k^2 \zeta \over 2(\rho +p)}\,.
\end{equation}
We divide by the frequency for a sound wave of wave-number $k$, take $v_s\approx 1/\sqrt{3}$, and shift to appropriate units,
obtaining, 
\begin{eqnarray}
\Gamma_{\rm damp}/(k v_s)\approx .28 \Bigr ( {k \over [1 \, Mpc]^{-1}}\Bigr )\Bigr ({ 10^{-6} \over
G }\Bigr )^4 \Bigr( {.2 eV \over T_0 }\Bigr )
\nonumber\\
\times H \Bigr ({m_\nu \over T_0}\Bigr )(1+k^2/k_0^2)^{-1}\, .~~~~~~~~~~~~~~
\label{decayrate}
\end{eqnarray}
where
\begin{equation}
{k_0\over [1\, {\rm Mpc]^{-1}}}= 30.3 \Bigr ({G \over  10^{-6}
}\Bigr )^4 \Bigr ({ T_0 \over .2\, eV }\Bigr )F\Bigr({m_\nu \over T_0  }\Bigr)\,.
\label{cut-off}
\end{equation}
From (\ref{decayrate}) and (\ref{cut-off}) we see that there can be strong damping in several
oscillations when the system is at the recombination temperature, for a neutrino mass that 
is of the order of 1 eV and coupling $G$ of the order $10^{-6}$. 
\section{Matter plus photon systems.} 
All of our development is applicable to the system of ionized hydrogen considered
by Weinberg, and we compare our result for this case to (\ref{weinberg})
in the low-frequency limit, $\zeta (\omega=0)$. This will provide both a check on our
pedestrian formalism, and possible application to ionized hydrogen systems at much higher densities 
and temperatures than prevail in the recombination region in cosmology.
In (\ref{bv0}), replacing the subscript $\phi$ by ${\gamma}$ and $\nu$ by ``mat", we use the adiabatic indices for the matter, $\gamma_{\rm mat}=5/3$, $\alpha_{\rm mat}
=2n_e$ and the specific heat, $c_V^{\rm (mat)}=3n_e$ where
$n_e$ is the density of the protons and the electrons. For the photons we use
the numbers from (\ref{params}) but with $\alpha$ and $C_V$ doubled, since there are
two polarization states. We obtain,
\begin{equation}
\zeta(0) ={ 16 a T^4 n_e^2 \over \Gamma(4 a T^3+3n_e)^2}\, ,
\label{ep}
\end{equation}
with $a=\pi^2/15$, the black-body constant in our units.
The Weinberg formula (\ref{weinberg}) yields the same result if we choose $\Gamma=
4 \tau_\gamma^{-1}$. We can check the relation between these parameters, in a system
in which photon emission and absorption (not scattering) provides the mechanism for
energy transfer, and in which the photon absorption rate is independent of energy, by
beginning from the Boltzmann equation for the photon distribution function $f(\omega)$,
\begin{eqnarray}
{\partial \over \partial \, t} f= \tau_\gamma^{-1}[ (1-e^{-\omega / T_m})f-e^{-\omega /T_m}]\, .
\end{eqnarray} 
We estimate the rate of energy transfer $\dot Q$ and from it the coefficient $\Gamma$ in (\ref{heat})
by inserting $f=(\exp[\omega/T_\gamma-1)^{-1}$, expanding to first order in $(T_\gamma-T_m)$,
multiplying by $\omega$ and integrating $d^3 k$. We obtain,
\begin{eqnarray} 
\Gamma={1 \over \tau_\gamma W(T_0)} \int_0^\infty  d \omega \, \omega^4
 { 1
\over [\exp(\omega/T_0)-1]}=3.83 \tau_\gamma^{-1}\, ,
\nonumber\\
\,
\label{tau-gamma}
\end{eqnarray}
where, 
\begin{equation}
W(T_0)=\int d\omega \, \omega^3 { 1\over \exp(\omega/T_0)-1}\, .
\end{equation}

We can directly apply (\ref{ep}) to determine
 bulk viscosity in regions in which it is dominated by the 
``free-free" photoemission and photoabsorption process, rather than by Compton scattering. This process is completely efficient in energy transfer, so that our previous caveats with respect
to an efficiency factor do not apply.  
The results will be most interesting in a system in which we have neither $n_{\rm e}>>n_\gamma$,
nor $n_{\rm e}<<n_\gamma$. We take the photo-absorption rate, $\Gamma_{\rm abs}$ from the Kramers formula, modified by the Gaunt factor correction \cite{iglesias};
\begin{eqnarray}
\Gamma_{\rm abs}(\omega)={16 \pi e^6 n_e^2 \over 3  m_e \omega^3}\sqrt {{2 \pi \over  m_e T}}
e^{\omega/(2T)}K_0 \Bigr ({\omega \over 2 T} \Bigr) \,,
\label{kramers} 
\end{eqnarray}
where $K_0$ is the modified Bessel function.

Defining the parameter to be used in this case for $\Gamma$ in (\ref{heat}) by $\Gamma_{ff}$,
we have, in place of (\ref{tau-gamma}),

\begin{eqnarray} 
\Gamma_{ff}={1\over W(T_0)}\int_0^\infty  d \omega \,\Gamma_{\rm abs} (\omega) 
{\omega^4 \over 
\exp(\omega/T_0)-1}
\nonumber\\
=114.\, e^6 m_e^{-3/2} T_0^{-7/2} n_e^2 \,.~~~~~~~~~~~~~~~~~~~~
\label{ff-gamma}
\end{eqnarray}

Taking $T/m_e$ as the efficiency factor for energy transfer in the Compton
process, we find that the bulk viscosity is determined by $\Gamma_{\rm ff}$,
rather than Compton scattering in all regions except those in which $n_e<<T^3$. 
In the latter regions, however it is much smaller than the shear viscosity
by virtue of the $n_e$ dependence of (\ref{ep}). 

\section{Discussion}
The principal result of this paper is the expression for the bulk viscosity of
the interacting gas of $\nu$'s and $\phi$'s. We took one neutrino flavor, and
an initial state with vanishing chemical potential, i. e. equal numbers
of $\nu$'s and $\bar \nu$'s (or equal numbers of right-handed and left-handed
neutrinos in the Majorana variation).

For the tightly coupled case with, say, $G\approx 10^{-5}$, we would find very
little dissipation in our models. For couplings of $G\approx 10^{-6}$, and a neutrino
mass of order of an eV we find relatively strong damping of a plane sound wave in the
medium. We note the following about this regime: 

1) In the simplest coupling schemes, at least, there would be a negligible
number of $\phi$'s produced prior to neutrino-matter decoupling. But for $G\approx 10^{-6}$
there would be production at a rate that serves to create an equilibrium distribution
during the period in which the neutrinos are still completely relativistic
and prior to photon decoupling.

2) In this domain, the shear viscosity and bulk viscosity will make comparable
contributions to the damping of a plane wave. However, when the shift is made
to a proper cosmological calculation, in which the shear viscosity is implicit
in the way the scattering rate enters the quadrupole terms, and the bulk velocity
damps monopoles, it is less clear to us what the relative contributions of bulk 
viscosity to the actual signal will be.

As we move into the region $G<10^{-6}$ the dissipation becomes greater, and
it appears that our effects would be an essential element of the analysis of the
perturbations. We note that the authors of ref \cite{raff} found that their
free-streaming criterion ruled out couplings down to a value $G\approx 10^{-7}$.

Finally, as will have been obvious, we made no attempt to include the complications
that would arise from three flavors of neutrinos with a mass splitting scheme to
agree with the data, let alone the possible complications of $\phi$ couplings that
carry anything other than the unit matrix in flavor space. When couplings that allow 
$\nu$ decays are included, the basic physics of our description remains unchanged,
but the numerology concerning the interesting domain of $G$'s becomes very different.

I thank Nicole Bell for discussions. This work was supported in part by NSF grant 
PHY-04559 18.
\section{Appendix}
We need the temperature response $\gamma_\nu -1$ to an adiabatic change in volume,
and the pressure response $\alpha_\nu$ to this temperature response, all of this
with the interactions between $\nu$ and $\phi$ that transfer energy between the
baths turned off. A complication is that in the true equilibrium state, in the
presence of the annihilation mechanism,
the number of $\nu$'s and $\bar \nu$'s (the latter, if we use Dirac $\nu$'s)
are not conserved. The average state, to which we apply our periodic disturbance,
thus retains the value $\mu=0$ for neutrino chemical potentials. But to conserve
neutrino (and anti-neutrino, if we use Dirac $\nu$'s) number in our periodic changes we need
to introduce an oscillating chemical potential perturbation $\delta \mu_\nu$ as well
as an oscillating value for $\delta V/V$. For the Dirac neutrino case the $\delta \mu_s$
are the {\it same} for neutrino and antineutrino (in contrast to an equilibrium
case, where they would be equal and opposite if there were zero net lepton number.)
We now have an energy equation,
\begin{eqnarray}
(P+\rho){\delta V\over V}=-{\partial \rho_\nu \over \partial T} \delta T-
{\partial \rho_\nu \over \partial \mu}\delta \mu_\nu\,,
\label{energy}
\end{eqnarray}
and a number conservation equation,
\begin{eqnarray}
{\delta V\over V}=-{1 \over n_\nu}\Bigr [{\partial n_\nu \over \partial T}\delta T +
{\partial n_\nu \over \partial \mu_\nu}\delta \mu \Bigr ]\,.
\label{number}
\end{eqnarray}
where $\mu_\nu$ is to be evaluated at zero after differentiation.
Defining,
\begin{eqnarray}
A=\Bigr[{1\over \rho_\nu+p_\nu}{\partial \rho_\nu \over \partial T}
-{1\over n}{\partial n\over \partial T}\Bigr ] \Bigr [ {{1\over n}{\partial n \over
\partial \mu}- {1 \over p+\rho_\nu}{\partial \rho_\nu \over \partial \mu_\nu} }\Bigr ]^{-1},
\nonumber\\\,
\end{eqnarray}
we have
\begin{equation}
\delta \mu_\nu=A \delta T ~,
\end {equation}
\begin{eqnarray}
\gamma_\nu-1=\Bigr [{\partial \rho_\nu \over \partial T}+A{\partial \rho_\nu \over \partial \mu_\nu}
 \Bigr ] [p_\nu+\rho_\nu]^{-1}~,
\nonumber\\
\,
\nonumber\\
\alpha_\nu={\partial p_\nu \over \partial T}.
\end{eqnarray}
 The quantities above are calculated from the expressions,
\begin{eqnarray}
n_\nu={1\over \pi^2}\int_0^\infty p^2 dp {1 \over 1+e^{(E-\mu_\nu)/T}}\,,
\end{eqnarray}
\begin{eqnarray}
\rho_\nu={1\over  \pi^2}\int_0^\infty p^2 dp {E \over 1+e^{(E-\mu_\nu)/T}}\,,
\end{eqnarray}
\begin{eqnarray}
p_\nu={1\over \pi^2}\int_0^\infty p^2 dp {p^2/E \over 1+e^{(E-\mu_\nu)/T}}\,,
\end{eqnarray}
where $E=\sqrt{p^2+m_\nu^2}$.
The parameters $\gamma_\nu-1$ , $\alpha_\nu$ are functions of the single variable,
$T/m_\nu$. In figs. 3 and 4. we show plots of the two parameters.

\begin {figure}[ht]
    \begin{center}
       \epsfxsize 2.75in
        \begin{tabular}{rc}
           \vbox{\hbox{
$\displaystyle{ \, { } }$
               \hskip -0.1in \null} 
} &
            \epsfbox{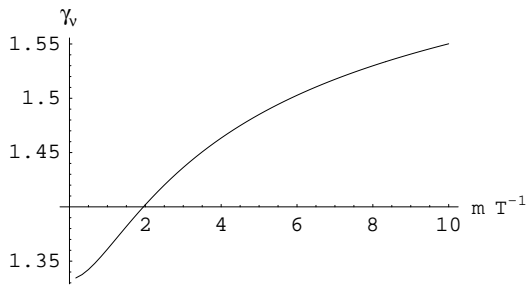} \\
            &
            \hbox{} \\
        \end{tabular}
    \end{center}
\label{fig.3}
\protect\caption
    {%
The index $\gamma_\nu(m_\nu/T)$.
 }
\end {figure}

\begin {figure}[ht]
    \begin{center}
       \epsfxsize 2.75in
        \begin{tabular}{rc}
           \vbox{\hbox{
$\displaystyle{ \, { } }$
               \hskip -0.1in \null} 
} &
            \epsfbox{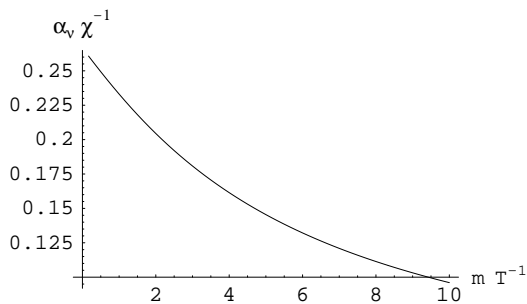} \\
            &
            \hbox{} \\
        \end{tabular}
    \end{center}
\label{fig.4}
\protect\caption
    {%
The function $\alpha_\nu(m_\nu/T)$ expressed as a multiple of the specific heat
$\chi=c_V$.
 }
\end {figure}
Note that $\gamma_\nu$
approaches $4/3$ at high temperatures, as expected. 
We note a borderline inconsistency in our procedures, above, in that
we will trust interactions that equilibrate the neutrino distribution thermally, such as $\nu-\nu$
scattering (from $\phi$ exchange) to do their job, while turning off (schematically)
the change in neutrino number and transfer of energy between the two seas, although all
of these effects are of the same order in $G$.

For the function $g(m_\nu /T)$ we take the expression given in eq(5) of ref. \cite{beacom}, for the $\nu$
annihilation rate, and multiply by the appropriate density ratio to obtain the estimate for the rate for
$\phi +\phi \rightarrow \nu + \nu$,
\begin{eqnarray}
\Gamma={G^4 \over 64\, \pi} {T \over m_\nu^3} \Bigr ( {m_\nu T \over 2 \pi}\Bigr )^{3/2}{n_\phi \over n_\nu} e^{-m_\nu /T}\,.
\end{eqnarray}

\end{document}